\documentclass[journal,11pt, onecolumn]{IEEEtran}
\usepackage{amsmath,graphicx}
\usepackage{graphicx}
\usepackage[noend]{algpseudocode}
\usepackage{indentfirst}
\usepackage{url}
\usepackage{algorithmicx,algorithm}
\usepackage{flushend}
\usepackage[noadjust]{cite}
\usepackage{subfigure}
\usepackage{multirow,tabularx}
\newcolumntype{C}{>{\centering\arraybackslash}X}
\newcolumntype{L}{>{\raggedright \arraybackslash}X}
\newcolumntype{R}{>{\raggedleft \arraybackslash}X}

\begin{document}
\title{Exploiting the Dual-Tree Complex Wavelet Transform  for Ship Wake Detection\\ in SAR Imagery}

\author{Wanli Ma, Alin Achim, Oktay Karakuş\thanks{This work was supported by the UK Engineering and Physical Sciences Research Council (EPSRC) under grant EP/R009260/1 (AssenSAR). }
        \thanks{Wanli Ma, Alin Achim and Oktay Karakuş are with the Visual Information Laboratory, University of Bristol, Bristol BS1 5DD, U.K. (e-mail: ix19797@bristol.ac.uk; alin.achim@bristol.ac.uk; o.karakus@bristol.ac.uk.)}
}

\maketitle
\begin{abstract}
In this paper, we analyse synthetic aperture radar (SAR) images of the sea surface using an inverse problem formulation whereby Radon domain information is enhanced in order to accurately detect ship wakes. This is achieved by promoting linear features in the images. For the inverse problem-solving stage, we propose a penalty function, which combines the dual-tree complex wavelet transform (DT-CWT) with the non-convex Cauchy penalty function. The solution to this inverse problem is based on the forward-backward (FB) splitting algorithm to obtain enhanced images in the Radon domain. The proposed method achieves the best results and leads to significant improvement in terms of various performance metrics, compared to state-of-the-art ship wake detection methods. The accuracy of detecting ship wakes in SAR images with different frequency bands and spatial resolution reaches more than 90\%, which clearly demonstrates an accuracy gain of 7\% compared to the second-best approach.
\end{abstract}
\section{Introduction}
\label{sec:intro}
\IEEEPARstart{T}{hanks} to the development of remote sensors technology, synthetic aperture radar (SAR) has become a widespread and effective imaging modality to acquire vital information for analysing and tracking vessels on the ocean surface. In SAR images of the ocean surface, a moving vessel generates different wake structures, which can be identified as: (i) \textit{turbulent wake}, (ii) \textit{Narrow V-wake}, and (iii) \textit{Kelvin wake}.
Due to speckle and the complex nature of the SAR imagery, the success of ship and wake detection remain limited, and new approaches have become necessary to address the aforementioned issues. In the literature, various works have suggested algorithms to leverage the information in SAR imagery for the purpose of wake detection. 

Considering the linear characteristics of ship wakes, their detection within SAR images is well suited to be performed in the Radon domain, which is the gold standard in line detection along with the Hough transform. Thus, Radon transform based methods have dominated the literature for ship wake detection solutions \cite{C8, C9,C10,C11,zilman2004speed,courmontagne2005improvement}.
Different from the previous approaches, Karakuş et. al. \cite{C12,C13} addressed ship wake detection as an inverse problem by expressing the observed SAR image in terms of the Radon domain information, and solution to the aforementioned inverse problem has been obtained by using sparse penalty functions such as the generalized minimax concave (GMC) penalty \cite{C24}. The approach in \cite{C12,C13} has then been implemented with a non-convex Cauchy based penalty function, with the ensuing algorithm guaranteed to convergence \cite{karakucs2019cauchy1}. This led to higher ship wake detection accuracy in \cite{C14, yang2020detection} compared to the GMC, $L_1$ and total variation (TV) penalties.

The dual-tree complex wavelet transform (DT-CWT) offers added benefits over the discrete wavelet transform (DWT). It calculates the complex transform of a signal using two separate DWT decompositions; one tree generates the real part of the transformation and another tree generates the imaginary part. DT-CWT is thus able to reduce the translation sensitivity and improve the directional selectivity of DWT. In the SAR imaging literature, owing to its advantages compared to the DWT, it has shown benefits in applications including despeckling \cite{ranjani2010dual, farhadiani2019hybrid}, sea ice detection \cite{gao2019sea}, and image enhancement \cite{nafornita2018multilook}.

In this paper, we propose a novel algorithm for detecting ship wakes, which is based on imaging inverse problems to enhance linear features in SAR imagery. We achieve this by combining the non-convex Cauchy based penalty function \cite{C14} with the DT-CWT. The proposed method, which benefits from the near shift-invariance property of the DT-CWT, along with the heavy-tailed nature of the Cauchy penalty function, is then compared to methods based on several state-of-the-art penalty functions, such as GMC \cite{C12}, $TV$, Cauchy \cite{C14}, and the method of Graziano et. al. \cite{C18}. All the corresponding minimisation problems are solved via the forward-backward (FB) proximal splitting algorithm.

The rest of the paper is organised as follows: we first present the theoretical preliminaries in Section \ref{sec:Theory}. The proposed method is discussed in Section \ref{sec:proposed}, whilst Section \ref{sec:results} presents the experimental analysis. Section \ref{sec:conc} concludes the paper with a brief summary and future work directions.
\section{THEORETICAL PRELIMINARIES}
\label{sec:Theory}
\subsection{Inverse Problem Formulation}
Since ship wakes are regarded as linear structures, the SAR image formation model can be defined in terms of the inverse Radon transform as \cite{C12}
\begin{equation}\label{equ:formation}
Y=\mathcal{C} X+N
\end{equation}
where $Y$ is the observed SAR image and $X(r, \theta)$ represents the information in the transform domain. Also, $\mathcal{C}$ refers to the inverse Radon transform, and $N$ is additive noise.

Under the assumption of an independent and identically distributed (iid) Gaussian noise, we express the data fidelity term (i.e. the likelihood) as
\begin{align}
    \mathcal{L}(Y, \mathcal{C}X) =  \|Y - \mathcal{C}X\|_2^2
\end{align}

Since recovering the object of interest $X$ from the observation $Y$ is an ill-posed inverse problem, we must scrutinise the prior information on $X$ to obtain a stable and unique reconstruction result. Having the prior knowledge $p(X)$, the problem of estimating $X$ from the observed SAR image $Y$ by using the signal model in (\ref{equ:formation}) turns into a minimisation problem as
\begin{align}
    \hat{X} = \arg\min_X \bigg\{ \|Y - \mathcal{C}X\|_2^2 - \log p(X) \bigg\}\label{equ:mini1}
\end{align}
where $\psi(X)=-\log p(X)$ is the penalty function, and thus the cost function takes the form $F(X) = \mathcal{L}(Y, \mathcal{C}X) + \psi(X)$.

\subsection{Dual-Tree Complex Wavelet Transform (DT-CWT)}

\begin{figure}[ht!]
\centering
\includegraphics[width=.9\linewidth]{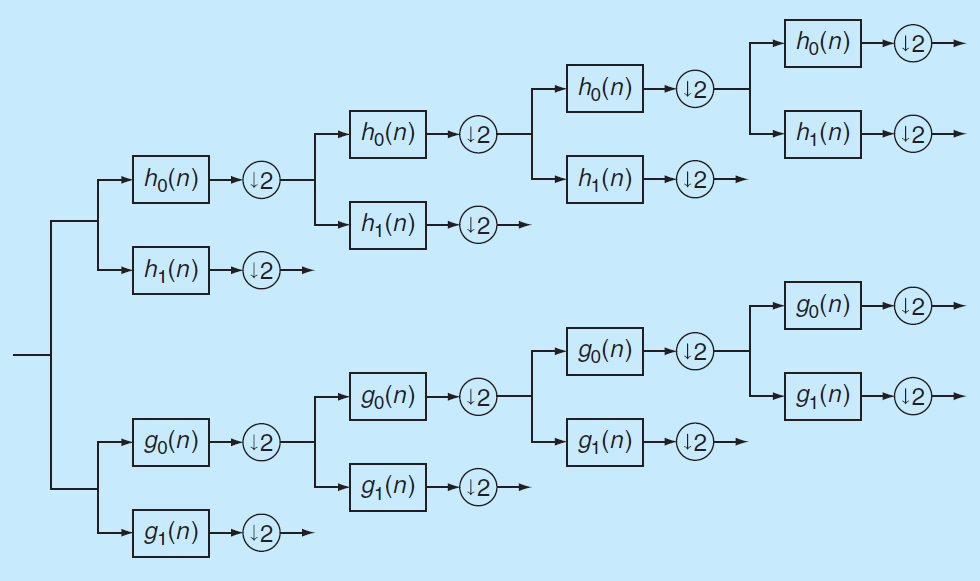}
\caption{Analysis filter bank for the DT-CWT \cite{C5}}
\label{fig:dtcwt}
\end{figure}

The DT-CWT is an efficient improvement of the classical DWT, which addresses the poor performance for complex and/or modulated signals, such as radar, speech, etc. As shown in \cite{C5}, two real DWTs constitute a DT-CWT, which has real and imaginary parts given by the first real DWT and the second real DWT, respectively. Figure \ref{fig:dtcwt} depicts the forward DT-CWT operation.

As long as we express two real DWTs as real ($\Psi_{h}(t)$) and imaginary ($\Psi_{g}(t)$) pair, the generic expression of the DT-CWT can be written as \cite{C5}
\begin{equation}
\Psi(\mathrm{t})=\Psi_{h}(t)+j \Psi_{g}(t)\label{equl:DTCWT}
\end{equation}
There is no data flow between $\Psi_{h}(t)$ and $\Psi_{g}(t)$, they thus can be calculated separately. 
Since in this paper we deal with images, a 2-D Dual-Tree CWT needs to be implemented as
\begin{equation}
\Psi(x, y)=\Psi(x) \Psi(y)\label{equl:2D DTCWT}
\end{equation}
Combining (\ref{equl:DTCWT}) and (\ref{equl:2D DTCWT}), we have
\begin{equation}
\Psi(x, y)=\left[\Psi_{h}(x)+j \Psi_{g}(x)\right]\left[\Psi_{h}(y)+j \Psi_{g}(y)\right]
\end{equation}


\section{THE PROPOSED METHOD}
\label{sec:proposed}
\subsection{Cauchy and DT-CWT Based Penalty Function}

The Cauchy distribution is a member of the symmetric $\alpha$-stable family, and the probability density function (pdf) of the Cauchy distribution is expressed as 
\begin{align}
    p(x) \propto\left[\frac{\gamma}{x^{2}+\gamma^{2}}\right]
\end{align}
where the parameter $\gamma$ is the scale parameter which determines the spread of the Cauchy distribution around zero. When replacing the prior information in (\ref{equ:mini1}) with the Cauchy prior, we obtain the following minimisation problem as \cite{karakucs2019cauchy1}
\begin{equation}
\hat{X}_{\text {Cauchy}}:=\arg \min _{\boldsymbol{X}}\left\{\|Y-\operatorname{CX}\|_{2}^{2}-\log \frac{\gamma}{X^{2}+\gamma^{2}}\right\}
\end{equation}
where the regularization term $-\log\frac{\gamma}{X^{2}+\gamma^{2}}$ is the non-convex Cauchy-based penalty function. 

In addition to its role in modelling the heavy-tailed noise in various signal and image processing applications, Cauchy based penalty function has rarely been used as the signal prior in the literature. Based on the derivation of its proximal operator in \cite{karakucs2019cauchy1}, the work in \cite{C14} proposed the Cauchy penalty function for addressing four SAR imaging inverse problems: super-resolution, image reconstruction, despeckling and ship wake detection. The Cauchy based penalty function has shown improved performance over various important penalty functions in \cite{C14} such as $L_1$, TV, and GMC. 

Although some efforts on ship wake detection  based on inverse problem solving have been done by employing penalty functions such as GMC \cite{C13}, and Cauchy \cite{C14}, both penalty functions are in need of further improvements mainly for better tackling noise artefacts, which sometimes induce false wake detections. DT-CWT appears a promising tool for efficiently enhancing the corresponding SAR image for the purpose of wake detection, thanks to its properties of reducing the translation sensitivity and improving the directional selectivity \cite{C5}. 
Hence, in order to benefit from its advantages, a second penalty function base on the DT-CWT is designed and added to the cost function
\begin{equation}\label{equ:propCost}
F(X)=\|Y-\mathcal{C} X\|_{2}^{2}-\log \frac{\gamma}{X^{2}+\gamma^{2}}+\lambda\|\mathcal{B} X\|_{1}
\end{equation}
where $\mathcal{B}$ is the forward DT-CWT operator, and $\|\cdot\|_{1}$ is $L_{1}$ norm which is a convex function. The inclusion of an additional term, $\|\mathcal{B} X\|_{1}$ involving the DT-CWT transform of $X$, contributes to the enhancement of the Radon transform $X$ of the SAR image $Y$, and promotes sparsity through the use of L1 norm. 

The solution for minimising the cost function in (\ref{equ:propCost}) can be based on proximal splitting algorithms such as the Forward-Backward (FB) \cite{C20} and Alternating Direction Method of Multipliers (ADMM) \cite{C21}. Thanks to its simpler implementation and success in image processing applications, in this work we prefer to utilise the FB algorithm. The FB splitting algorithm first necessitates to split the cost function into two functions as
\begin{align}
   \hat{X}_{\text{Cauchy-DTCWT}} = \arg \min_{X}\bigg\{ F(X)=g(X)+h(X)\bigg\}.
\end{align}

To this end, having a convex data fidelity term and the differentiable Cauchy penalty function, we choose functions $g(\cdot)$ and $h(\cdot)$
\begin{align}
   g(X) &= \|Y-\mathcal{C} X\|_{2}^{2}-\log \frac{\gamma}{X^{2}+\gamma^{2}},\\
   h(X) &= \lambda\|\mathcal{B} X\|_{1}.
\end{align}

The splitting of the cost function mentioned above only requires the proximal operator of $h(X)$. 
Thus, for any $\mu>0$, the iterative FB solution is obtained by
\begin{equation}
X^{(\ell+1)}=\operatorname{prox}_{\mu h}\left(X^{(\ell)}-\mu \nabla g\left(X^{(\ell)}\right)\right)
\end{equation}
where the gradient of the function $g$ becomes 
\begin{align}
   \nabla g(X)=\mathcal{C}^{T}(\mathcal{C} X-Y)+\frac{2 X}{\gamma^{2}+X^{2}}. 
\end{align}

The proximal operator of the $L_{1}$ norm-based penalty functions is the well-known soft thresholding function
\begin{align}
    \operatorname{prox}_{\mu h}(X)=\max \left(0,1-\frac{\mu \lambda}{B X}\right).
\end{align}

Consequently, the iterative solution to the proposed minimisation problem in (\ref{equ:propCost}) is as shown in Algorithm 1 below. 
We set the maximum number of iterations (maxIter) to 500. The error term $\varepsilon^{(i)}$ is set to 0.005, which can be calculated as: $\epsilon^{(i)}=\frac{\left\|X^{(i)}-X^{(i-1)}\right\|}{\left\|X^{(i-1)}\right\|}$.

\begin{algorithm}[h]
\caption{FB Algorithm for the proposed penalty function} 
\hspace*{0.02in} {\bf Input:} 
SAR imagery $Y$ and $\mu, \gamma, \lambda$ \\
\hspace*{0.02in} {\bf Output:} 
Radon image X\\
\hspace*{0.02in} {\bf Set:}
$\boldsymbol{i} = 0$ and $\boldsymbol{X}^{(0)}=\{\boldsymbol{0}\}$
\begin{algorithmic}[1]
\State  {\bf do:}
\State $Z^{(i)}= X^{(i)}-\mu\nabla(g(X^{(i)})).$
\State $\mathbf{W}=\operatorname{soft}\left\{\mathbf{B}\left(\mathbf{Z}^{(i)}\right), \boldsymbol{\mu} * \lambda\right\}$
\State $X^{(i+1)}={\mathcal{B}}^{-1}(\mathbf{W})$
\State $i++$\\
\hspace*{0.00in} {\bf while:}
$\varepsilon^{(i)}>10^{-3}$ or $i<$ maxIter
\end{algorithmic}
\end{algorithm}
\subsection{Ship Wake Detection}
\begin{figure}[ht!]
\centering
\includegraphics[width=.65\linewidth]{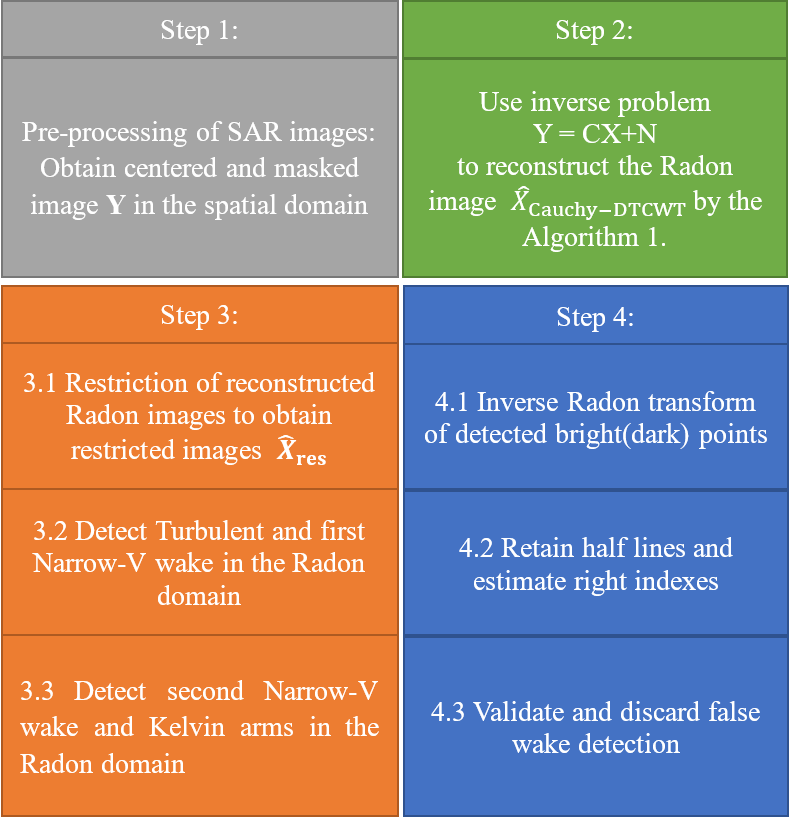}
\caption{Ship wake detection Method.}
\label{fig:steps}
\end{figure}

As soon as we have the reconstructed Radon information from the optimisation procedure detailed above, in order to detect the wake structures
, we use the same detection method as previously proposed in \cite{C12}. The computational steps of the whole ship wake detection algorithm in this paper are presented in Figure \ref{fig:steps}. For further details please see \cite{C12}.

For the validation in Step 4, we use the merit index $\mathrm{F}_{\mathrm{t}}^{\mathrm{I}}=\frac{\overline{\mathrm{I}}_{\mathrm{t}}}{\overline{\mathrm{I}}}-1$ \cite{C18}, in which ${\overline{\mathrm{I}}_{\mathrm{t}}}$ is the mean value along the half wake, and ${\overline{\mathrm{I}}}$ is the average intensity of the SAR image. For the turbulent wake validation, $\mathrm{F}_{\mathrm{t}}^{\mathrm{I}}$ is allowed to be a negative value. For other wake structures $\mathrm{F}_{\mathrm{t}}^{\mathrm{I}}$ must be greater than zero. We also set a margin of 10\% to reduce the possibility of false detections, thus for the Narrow-V and Kelvin wakes, $\mathrm{F}_{\mathrm{t}}^{\mathrm{I}}$ needs to be greater than 0.1.

\section{Experimental results}
\label{sec:results}
\subsection{SAR Image Data}
In this paper, 22 real SAR images from five different satellite platforms including TerraSAR-X, ALOS2, COSMO-SkyMed, Sentinel-1, and ICEYE were used. Seven of them are X-band SAR images from TerraSAR-X,  with a resolution of 3$\times$3 m. We also tested four L-band SAR images from ALOS2, which also have a resolution of 3$\times$3 m. The COSMO-SkyMed (5 images) and ICEYE (1 image) also provide X-band SAR images. The ICEYE \cite{C22} image utilised in this work is an HH-polarised, Stripmap product, which has 3-m resolution for both azimuth and range directions. Sentinel-1 (5 images) utilises C-band frequency and the resolution of the images is 10$\times$10 m and hence includes less details than the images from the other satellite platforms used. In Table \ref{tab:addlabel}, each SAR image is shown with their visible and invisible wakes (1 and 0 values, respectively) corresponding to a visual analysis.

\begin{table}[htbp]
\caption {Wake information of the SAR imagery data sets. T: Turbulent, N1 \& N2: Narrow V-wake, K1 \& K2: Kelvin wake.} 
\centering
\begin{tabularx}{0.76\linewidth}{@{} Cp{5cm}CCCCC @{}}
\hline
\textbf{Image} & \textbf{Image}    & \multicolumn{5}{c}{\textbf{Wake Types}} \\
\textbf{} &  \textbf{Source}  & \textit{T}   & \textit{N1} & \textit{N2}  & \textit{K1} & \textit{K2} \\
 \hline
\textbf{1} & TerraSAR-X      & 1     & 1     & 0     & 0     & 1 \\
\textbf{2} & TerraSAR-X      & 1     & 1     & 0     & 1     & 0 \\
\textbf{3} & ALOS2       & 1     & 1     & 0     & 0     & 0 \\
\textbf{4} & ALOS2       & 1     & 1     & 0     & 1     & 0 \\
\textbf{5} & COSMO-SkyMed      & 1     & 1     & 0     & 0     & 0 \\
\textbf{6} & COSMO-SkyMed      & 1     & 1     & 0     & 1     & 0 \\
\textbf{7} & COSMO-SkyMed      & 1     & 1     & 0     & 1     & 0 \\
\textbf{8} & COSMO-SkyMed      & 1     & 1     & 0     & 0     & 0 \\
\textbf{9} & Sentinel-1       & 1     & 1     & 0     & 0     & 0 \\
\textbf{10} & Sentinel-1       & 1     & 1     & 0     & 0     & 0 \\
\textbf{11} & TerraSAR-X      & 1     & 1     & 0     & 1     & 0 \\
\textbf{12} & TerraSAR-X      & 1     & 1     & 0     & 1     & 0 \\
\textbf{13} & TerraSAR-X      & 1     & 1     & 0     & 1     & 0 \\
\textbf{14} & TerraSAR-X      & 1     & 1     & 0     & 1     & 0 \\
\textbf{15} & COSMO-SkyMed      & 1     & 1     & 0     & 0     & 0 \\
\textbf{16} & ALOS2       & 1     & 1     & 0     & 0     & 0 \\
\textbf{17} & ALOS2       & 1     & 1     & 0     & 0     & 0 \\
\textbf{18} & Sentinel-1       & 1     & 1     & 0     & 0     & 0 \\
\textbf{19} & Sentinel-1       & 1     & 1     & 0     & 0     & 0 \\
\textbf{20} & Sentinel-1       & 1     & 1     & 0     & 0     & 0 \\
\textbf{21} & ICEYE      & 1     & 1     & 0     & 1     & 0 \\
\textbf{22} & TerraSAR-X      & 1     & 1     & 0     & 0     & 0 \\
\hline
\end{tabularx}%
\label{tab:addlabel}%
\end{table}

\subsection{Performance Analysis}
We compared the proposed method 
to four state-of-the-art methods: (1) the detection algorithm proposed by Graziano et. al. \cite{C18}, (2) the method based on GMC penalty function \cite{C12}, (3) total variation (TV) norm-based method, (4) method using a non-convex Cauchy penalty function \cite{C14}.
All 22 images mentioned above were centred on the ship and masked after appropriate pre-processing stages. For each image, ship wake structures are classified into three categories: (1) Turbulent Wake, (2) narrow-V Wake (2 arms) and (3) Kelvin wake (2 arms) \cite{C16}, and the detection performance metrics are calculated depending on correct detection/discard of these wakes.
The ship wake detection procedure was implemented by using the AssenSAR Wake Detector Matlab software \cite{C25} available as open-source, and for the DT-CWT implementation we used the sample functions published in \cite{C26}. 
The detailed performance analysis for all images is shown in Table \ref{tab:ROC} and Table \ref{tab:perf} with the best results in bold. 

\begin{table}[t]
\caption {Detection performance for all SAR images in terms of the receiver operating characteristic (ROC) metrics} 
\centering
\begin{tabularx}{0.76\linewidth}{@{}p{5cm}CCCC @{}}
\hline
\textbf{Methods} & TP &  TN  & FP    & FN      \\
\hline
Graziano et. al. \cite{C18}&	32.73$\%$&	38.18$\%$&	25.45$\%$&	3.64$\%$\\
GMC \cite{C13}&	\textbf{43.64$\%$}&	34.55$\%$&	18.18$\%$&	3.64$\%$\\
TV&	38.18$\%$&	33.64$\%$&	26.36$\%$&	\textbf{1.82$\%$}\\
Cauchy \cite{C14}	&37.27$\%$&	47.27$\%$&	10.00$\%$&	6.36$\%$\\
\newcommand{\tabincell}[2]{\begin{tabular}{@{}#1@{}}#2\end{tabular}}
Proposed &	41.82$\%$&	\textbf{48.18$\%$}&	\textbf{5.45$\%$}&	5.45$\%$\\
\hline
\end{tabularx}%
\label{tab:ROC}\vspace{1cm}%

\caption{Detection performance of different methods for all SAR images} 
\centering
\begin{tabularx}{0.76\linewidth}{@{}p{5cm}CCCC @{}}
\hline
\textbf{Methods} & \% Accuracy&	$F_{1}$ &	LR+ &	Youden's J    \\
\hline
Graziano et. al. \cite{C18}&70.91$\%$&	0.69&	2.25&	0.50\\
GMC \cite{C13}&77.27$\%$&	0.80&	2.68&	0.58\\
TV&	70.91$\%$&	0.73&	2.17&	0.52\\
Cauchy \cite{C14}	&84.55$\%$&	0.82&	4.89&	0.68\\
\newcommand{\tabincell}[2]{\begin{tabular}{@{}#1@{}}#2\end{tabular}}
Proposed &	\textbf{90.91$\%$}&	\textbf{0.88}&	\textbf{8.70}&	\textbf{0.78}\\
\hline
\end{tabularx}%
\label{tab:perf}%
\end{table}



In terms of  True Positive (TP) values in Table \ref{tab:ROC}, the GMC based method shows the best performance with 43.64$\%$. The TP of the proposed method is 41.82$\%$ which is only marginally lower than that of GMC, and higher than that of Graziano et. al., TV, and Cauchy. As shown in Table \ref{tab:ROC} Cauchy and DT-CWT based prior led to superior performance in terms of the ROC metrics corresponding to True Negative (TN) and False Positive (FP), which are 48.18$\%$ and 5.45$\%$, respectively. 

We also computed four quantitative metrics for evaluating the detection performance, which are presented in Table \ref{tab:perf}. In particular, for all four performance metrics, the proposed method achieves the best results with 90.91\% detection accuracy. Compared to the results of the other four state-of-the-art methods, the proposed method outperforms all by achieving around 6.5$\%$ accuracy gain compared to the Cauchy \cite{C14}, and approximately 20$\%$ compared to Graziano et. al. \cite{C18} and TV. 

\begin{figure}[t]
\centering
\subfigure[SAR image]{
\includegraphics[width=0.32\linewidth]{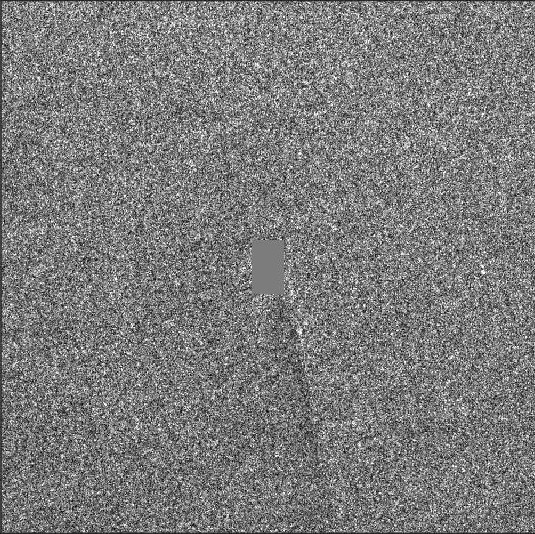}}
\subfigure[Graziano et. al. \cite{C18}]{
\includegraphics[width=0.32\linewidth]{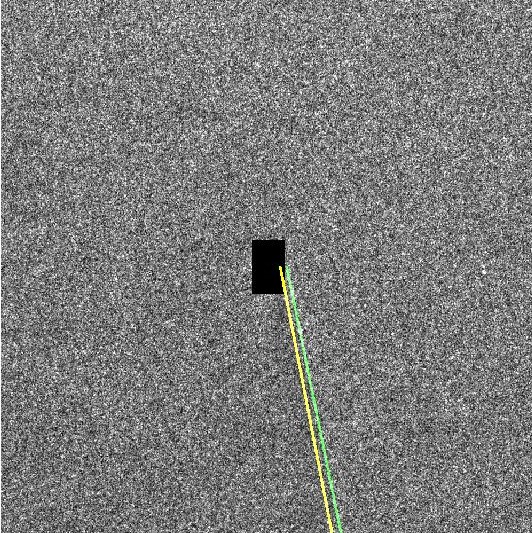}
}
\subfigure[TV]{
\includegraphics[width=0.31\linewidth]{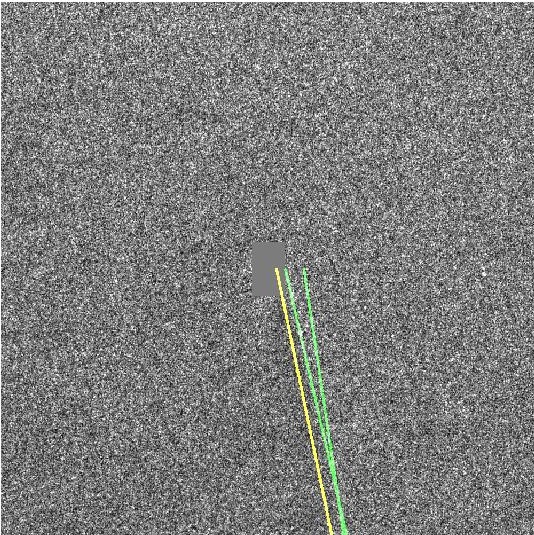}
}
\subfigure[GMC \cite{C13}]{
\includegraphics[width=0.32\linewidth]{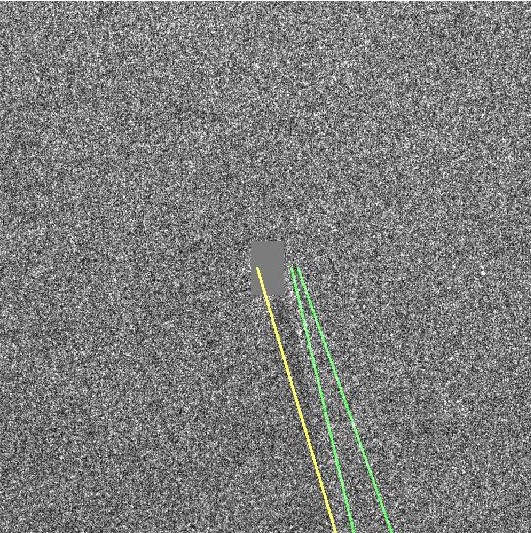}
}
\subfigure[Cauchy \cite{C14}]{
\includegraphics[width=0.32\linewidth]{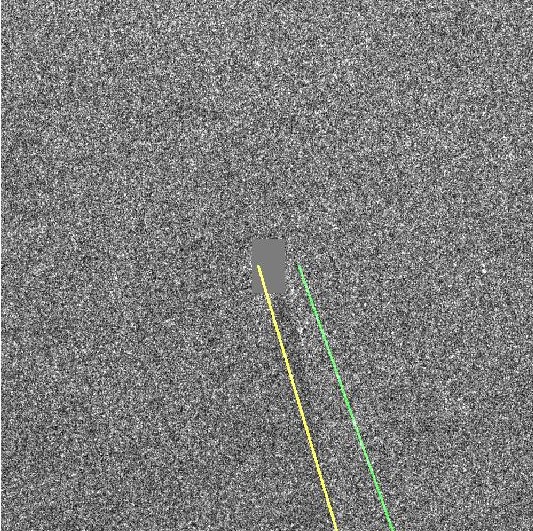}
}
\subfigure[Proposed method]{
\includegraphics[width=0.32\linewidth]{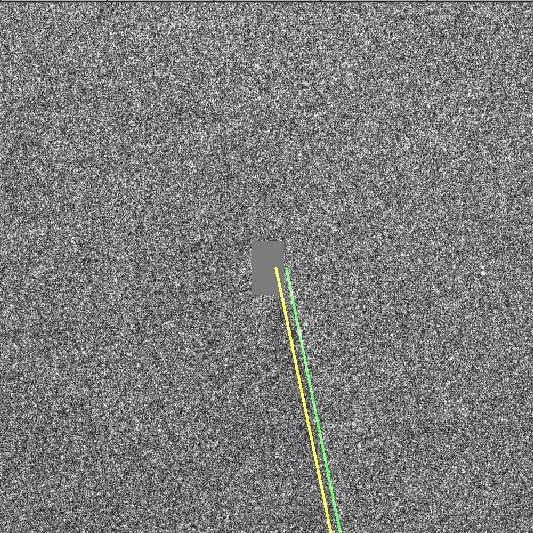}
}
\caption{Ship wake detection results for SAR Image 9. Yellow and green lines refer to the detected turbulent and narrow-V wakes, respectively.}
\label{fig:image9}
\end{figure}

In order to illustrate the numerical performance analysis given above, for visual analysis we also present detection results corresponding to an example SAR image, which is depicted in Figure \ref{fig:image9}. When examining Figure \ref{fig:image9}, Graziano et. al. \cite{C18} and the proposed method show remarkable detection results for the example considered. However, TV, GMC and Cauchy based approaches have relatively lower accuracy results  (in particular for the detection of narrow-V wakes).

Considering the proposed method in this paper is an improvement of the method in \cite{C14}, we can deduce that the addition of the DT-CWT based penalty to the Cauchy-based penalty term led to a significant performance increase, in particular for reducing false wake detections (FP values in Table \ref{tab:ROC}). Since false detections correspond to noisy peak values in the reconstructed Radon images, the DT-CWT penalty addition helps to reduce the noise in Radon domain and turn these false detections into correct detections (TP values in Table \ref{tab:ROC}).

\section{Conclusions}
\label{sec:conc}

In this paper, we proposed a novel regularisation strategy, which combines the non-convex Cauchy based penalty function with an additional term based on the DT-CWT. A proximal splitting optimisation algorithm was used to solve the inverse problem and to enhance the information in the Radon domain. 

The results of the performance analysis, which was based on 22 SAR images from five different satellite platforms, demonstrates that the proposed approach detects ship wakes with a high accuracy of around 91$\%$. It is obvious that the proposed penalty function benefits from the advantages of DT-CWT (improving the translation invariance and the directional feature selectivity \cite{C5}) and removes thus noisy artefacts from the reconstructed images. The proposed method achieves better detection performance than existing inverse problem based ship wake detection methods. On the other hand, the proposed method still offers potential for further improvement (e.g. to achieve higher TP) and developing more approaches, based on other types of sparse representations, is part of our current endeavours.

\section{Acknowledgment}
We are grateful to the UK Satellite Applications Catapult and to OceanMind Ltd for providing the COSMO-SkyMed  and the Sentinel-1 data sets, respectively. 


\bibliographystyle{IEEEbib}
\bibliography{refs}

\begin{thebibliography}{10}

\bibitem{C8}
Lesley~M Murphy,
\newblock ``Linear feature detection and enhancement in noisy images via the
  {Radon} transform,''
\newblock {\em Pattern recognition letters}, vol. 4, no. 4, pp. 279--284, 1986.

\bibitem{C9}
Maria~Daniela Graziano, Marco D’Errico, and Giancarlo Rufino,
\newblock ``Ship heading and velocity analysis by wake detection in {SAR}
  images,''
\newblock {\em Acta astronautica}, vol. 128, pp. 72--82, 2016.

\bibitem{C10}
James~KE Tunaley,
\newblock ``The estimation of ship velocity from {SAR} imagery,''
\newblock in {\em IGARSS 2003. 2003 IEEE International Geoscience and Remote
  Sensing Symposium. Proceedings (IEEE Cat. No. 03CH37477)}. IEEE, 2003,
  vol.~1, pp. 191--193.

\bibitem{C11}
Maria~T Rey, James~K Tunaley, JT~Folinsbee, PAUL~A Jahans, JA~Dixon, and
  Malcolm~R Vant,
\newblock ``Application of {Radon} transform techniques to wake detection in
  {Seasat-A} {SAR} images,''
\newblock {\em IEEE Transactions on Geoscience and Remote Sensing}, vol. 28,
  no. 4, pp. 553--560, 1990.

\bibitem{zilman2004speed}
Gregory Zilman, Anatoli Zapolski, and Moshe Marom,
\newblock ``The speed and beam of a ship from its wake's {SAR} images,''
\newblock {\em IEEE Transactions on Geoscience and Remote Sensing}, vol. 42,
  no. 10, pp. 2335--2343, 2004.

\bibitem{courmontagne2005improvement}
Ph~Courmontagne,
\newblock ``An improvement of ship wake detection based on the {Radon}
  transform,''
\newblock {\em Signal Processing}, vol. 85, no. 8, pp. 1634--1654, 2005.

\bibitem{C12}
Oktay Karakuş, Igor Rizaev, and Alin Achim,
\newblock ``Ship wake detection in {SAR} images via sparse regularization,''
\newblock {\em IEEE Transactions on Geoscience and Remote Sensing}, vol. 58,
  no. 3, pp. 1665--1677, March 2020.

\bibitem{C13}
Oktay Karaku{\c{s}} and Alin Achim,
\newblock ``Ship wake detection in {X-band SAR} images using sparse {GMC}
  regularization,''
\newblock in {\em ICASSP 2019-2019 IEEE International Conference on Acoustics,
  Speech and Signal Processing (ICASSP)}. IEEE, 2019, pp. 2182--2186.

\bibitem{C24}
Ivan Selesnick,
\newblock ``Sparse regularization via convex analysis,''
\newblock {\em IEEE Transactions on Signal Processing}, vol. 65, no. 17, pp.
  4481--4494, 2017.

\bibitem{karakucs2019cauchy1}
Oktay Karaku{\c{s}}, Perla Mayo, and Alin Achim,
\newblock ``Convergence guarantees for non-convex optimisation with
  {Cauchy}-based penalties,''
\newblock {\em IEEE Transactions on Signal Processing}, vol. 68, pp.
  6159--6170, 2020.

\bibitem{C14}
Oktay Karaku{\c{s}} and Alin Achim,
\newblock ``On solving {SAR} imaging inverse problems using nonconvex
  regularization with a {Cauchy}-based penalty,''
\newblock {\em IEEE Transactions on Geoscience and Remote Sensing}, pp. 1--13,
  2020.

\bibitem{yang2020detection}
Tianqi Yang, Oktay Karakuş, and Alin Achim,
\newblock ``Detection of ship wakes in {SAR} imagery using {Cauchy}
  regularisation,''
\newblock in {\em 2020 IEEE International Conference on Image Processing
  (ICIP)}, 2020, pp. 3473--3477.

\bibitem{ranjani2010dual}
J~Jennifer Ranjani and SJ~Thiruvengadam,
\newblock ``Dual-tree complex wavelet transform based {SAR} despeckling using
  interscale dependence,''
\newblock {\em IEEE Transactions on geoscience and remote sensing}, vol. 48,
  no. 6, pp. 2723--2731, 2010.

\bibitem{farhadiani2019hybrid}
Ramin Farhadiani, Saeid Homayouni, and Abdolreza Safari,
\newblock ``Hybrid {SAR} speckle reduction using complex wavelet shrinkage and
  non-local {PCA-based} filtering,''
\newblock {\em IEEE Journal of Selected Topics in Applied Earth Observations
  and Remote Sensing}, vol. 12, no. 5, pp. 1489--1496, 2019.

\bibitem{gao2019sea}
Feng Gao, Xiao Wang, Yunhao Gao, Junyu Dong, and Shengke Wang,
\newblock ``Sea ice change detection in {SAR} images based on
  convolutional-wavelet neural networks,''
\newblock {\em IEEE Geoscience and Remote Sensing Letters}, vol. 16, no. 8, pp.
  1240--1244, 2019.

\bibitem{nafornita2018multilook}
Corina Nafornita, Alexandru Isar, and Teodor Dehelean,
\newblock ``Multilook {SAR} image enhancement using the dual tree complex
  wavelet transform,''
\newblock in {\em 2018 International Conference on Communications (COMM)}.
  IEEE, 2018, pp. 151--156.

\bibitem{C18}
Maria~Daniela Graziano, Marco D’Errico, and Giancarlo Rufino,
\newblock ``Wake component detection in {X-band SAR} images for ship heading
  and velocity estimation,''
\newblock {\em Remote Sensing}, vol. 8, no. 6, pp. 498, 2016.

\bibitem{C5}
Ivan~W Selesnick, Richard~G Baraniuk, and Nick~C Kingsbury,
\newblock ``The dual-tree complex wavelet transform,''
\newblock {\em IEEE signal processing magazine}, vol. 22, no. 6, pp. 123--151,
  2005.

\bibitem{C20}
Numerical-tours.com,
\newblock ``Forward-backward proximal splitting,''
  \url{http://www.numerical-tours.com/matlab/optim_4_fb/} Accessed: 07- Sep-
  2020.

\bibitem{C21}
Stephen Boyd, Neal Parikh, and Eric Chu,
\newblock {\em Distributed optimization and statistical learning via the
  alternating direction method of multipliers},
\newblock Now Publishers Inc, 2011.

\bibitem{C22}
ICEYE,
\newblock ``Download: {SAR} datasets,''
  \url{https://www.iceye.com/downloads/datasets/} Accessed: 09- Sep- 2020.

\bibitem{C16}
W.~G. Pichel, P.~C. Colon, C.~C. Wackerman, and K.~S. Friedman,
\newblock ``Ship and wake detection,'' \url{http://www.sarusersmanual.com/}
  Accessed: 06- Sep- 2020.

\bibitem{C25}
Oktay Karaku{\c{s}} and Alin Achim,
\newblock ``{AssenSAR} wake detector,'' University of Bristol Data Repository,
\newblock \url{https://doi.org/10.5523/bris.f2q4t5pqlix62sv5ntvq51yjy}.

\bibitem{C26}
``Wavelet software at {Brooklyn Poly},'' Eeweb.poly.edu,
\newblock \url{http://eeweb.poly.edu/iselesni/WaveletSoftware}.

\end{thebibliography}
\end{document}